\documentclass[preprint,preprintnumbers,amsmath,amssymb]{revtex4}

\usepackage{graphicx}% Include figure files
\usepackage{dcolumn}% Align table columns on decimal point
\usepackage{bm}% bold math
\usepackage{psfrag}
\usepackage{epsfig}

\newcommand{\mb}{\mbox}
\newcommand{\beq}{\begin{equation}}
\newcommand{\eeq}{\end{equation}}
\newcommand{\lb}{\label}
\newcommand{\beqa}{\begin{eqnarray}}
\newcommand{\eeqa}{\end{eqnarray}}
\newcommand{\nnb}{\nonumber}
\newcommand{\la}{\langle}
\newcommand{\ra}{\rangle}

\begin{document}

%\preprint{APS/123-QED}

\title{Onsager's Inequality, the Landau-Feynman Ansatz and Superfluidity}

\author{Walter F. Wreszinski}
\email{wreszins@fma.if.usp.br}
\author{Milton A. da Silva Jr.}
\email{milton@fma.if.usp.br}
\affiliation{Instituto de F\'{\i}sica, Universidade de S\~ao Paulo\\
CP 66318, 05315-970 S\~ao Paulo, Brazil}

\date{\today}

\begin{abstract}

We revisit an inequality due to Onsager, which states that the
(quantum) liquid structure factor has an upper bound of the form
$(const.)\times|\vec{k}|$, for not too large modulus of the wave
vector $\vec{k}$. This inequality implies the validity of the Landau
criterion in the theory of superfluidity with a definite, nonzero
critical velocity. We prove an auxiliary proposition for general Bose
systems, together with which we arrive at a rigorous proof of the
inequality for one of the very few soluble examples of an interacting
Bose fluid, Girardeau's model. The latter proof demonstrates the
importance of the thermodynamic limit of the structure factor, which
must be taken initially at $\vec{k}\neq\vec{0}$. It also substantiates
very well the heuristic density functional arguments, which are also
shown to hold exactly in the limit of large wave-lengths. We also
briefly discuss which features of the proof may be present in higher
dimensions, as well as some open problems related to superfluidity of
trapped gases.

\end{abstract}

%\pacs{67.40.-w,67.40.Bz,67.40.Db,67.40.Kh}
\maketitle

\section{Introduction}

The properties of the Fourier transform of the pair (or
density-density) correlation - in particular its extrapolation to zero
angle, and the corresponding relation to the isothermal
compressibility - have been thoroughly studied in connection with the
critical behaviour of classical fluids (see M. E. Fisher's early
review \cite{fisher1}). The situation is very different regarding the
corresponding quantity - which we shall call the liquid structure
factor - for quantum Bose fluids \cite{price}. Even for the
impenetrable Bose gas - a prototype of one-dimensional systems with
on-site repulsion - very few results on correlation functions exist,
notably on the Fourier transform of the one-particle density matrix, a
mathematical tour-de-force by Lenard \cite{lenard}. Some results on
higher-order correlations exists as well \cite{forrester}, in
connection with the so-called Fisher-Hartwig conjecture
\cite{fisher3}, but they are restricted to Dirichlet or Neumann
boundary conditions (b.c.).   

The quantum liquid structure factor plays a crucial role in the
Feynman variational Ansatz \cite{feynman1}, but periodic b.c. are
essential in this context, because the corresponding wave-function
must have definite momentum. 

In this paper we present a rigorous proof of Onsager's inequality
\cite{price} for the liquid structure factor in the
one-dimensional impenetrable Bose gas (Girardeau's model
\cite{girardeau}). This inequality implies the validity of Landau's
criterion of superfluidity \cite{zagrebnov} with a definite, nonzero
critical velocity. Some features of the proof hold - others are
expected to hold - in higher dimensions, as discussed in the
conclusion. 

\section{Onsager's Inequality, the Landau Criterion and the
Feynman Ansatz. Summary}

Several criteria for superfluidity exist in the literature
\cite{hohenberg}. By one of the standard criteria \cite{hohenberg}, the
free Bose gas is a superfluid, while by the Landau criterion (see,
e.g. \cite{zagrebnov}), it is not. Landau postulated a spectrum of
elementary excitations (see, e.g., Fig. 13.11 of \cite{huang}), which
leads, by a well-known argument, to the startling property of
\underline{superfluidity} of the Bose fluid, i.e., the fact that under
certain conditions its viscosity (e.g., of liquid {\bf He-II}) vanishes,
i.e., the liquid is capable of flowing without dissipation through
very thin capillaries, as long as the velocity $\vec{v}$ of the liquid
has an absolute value below a certain critical speed $v_c$:

\beq
\lb{1}
|\vec{v}|\leq v_c.
\eeq

In general, $v_c$ increases as the diameter of the capillary
decreases. See \cite{huang} or \cite{martin} as excellent introductory
texts.

The Hamiltonian of a system of $N$ particles in a cube $\Lambda$ of
side $L$, upon which periodic b.c. are imposed, may be written (we use
units such that $\hbar=m=1$, where $m$ is the mass of the particles):

\beq
\lb{2}
H_{\Lambda} = -\frac{1}{2}\sum_{j=1}^{N}\Delta_{j} +
V(\vec{x}_1,\ldots,\vec{x}_N).
\eeq 

\noindent The momentum operator may be written

\beq
\lb{3}
\vec{P}_{\Lambda}= -\text{i}\sum_{j=1}^{N}\nabla_{j}.
\eeq

\noindent Above, 

\beq
\lb{4}
V=\sum_{1\leq i<j \leq N}\Phi(\vec{x}_{i} - \vec{x}_{j}),
\eeq

\noindent where $\Phi$ is the interparticle potential, which we assume
to satisfy the conditions appropriate to existence of the
thermodynamic limit \cite{fisher2}. Let

\begin{subequations}
\lb{5}
\beq
\lb{5a}
U_{\vec{v}}\equiv\mb{e}^{\text{i}\vec{v}\cdot(\vec{x}_1+\cdots+\vec{x}_N)}
\eeq 

\noindent be the unitary operator which implements Galilean
transformations. The operator implementing unitary
Galilean transformation has actually the form 

\beq
\lb{5b}
U^t_{\vec{v}}=\mb{e}^{\text{i}\vec{v}\cdot[(\vec{x}_1+
\cdots+\vec{x}_N)-t\vec{P}_{\Lambda}]}.
\eeq
\end{subequations}

\noindent However we may restrict ourselves to $t=0$ in (\ref{5b}),
and use (\ref{5a}) instead, just for the purpose of obtaining the
transformed Hamiltonian (energy), which indeed yields the correct
formula (\ref{7}). The reason for the latter is that the commutator of
the generator of $U_{\vec{v}}$ in (\ref{5b}) with $H_\Lambda$,
independs of $t$, because 

\beq
\lb{6}
[H_\Lambda,\vec{P}_\Lambda]=0.
\eeq

\noindent The operators are defined on the Hilbert Space 
$\mathcal{H}_{\Lambda}=L_{\text{sym}}^2 (\Lambda^N)$, of symmetric
square integrable wave-functions over the $N$-fold Cartesian product
of the cube $\Lambda$ with itself, satisfying periodic b.c. on
$\Lambda$. We shall disregard the spin variable. We have

\beq
\lb{7}
\widetilde{H}_{\Lambda}^{\vec{v}}\equiv
U_{\vec{v}}^{\ast}H_{\Lambda}U_{\vec{v}} = H_{\Lambda} +
\vec{v}\cdot\vec{P}_{\Lambda} + \frac{1}{2}N\vec{v}^2.
\eeq

Suppose the fluid moves with constant velocity $\vec{v}$ with respect
to some fixed inertial frame $F$. In the rest frame of the fluid the
Hamiltonian is $H_{\Lambda}$, and by (\ref{6})
$\widetilde{H}_{\Lambda}^{\vec{v}}$ is the Hamiltonian with respect to the
frame $F$. In the rest frame of the fluid, it posesses a unique ground
state $\Omega_{\Lambda}$, with energy $E_{0}(N,L)$ say, and zero
momentum:

\begin{subequations}
\lb{8}
\beqa
\lb{8a}
H_{\Lambda}\Omega_{\Lambda} &=& E_{0}(N,L)\Omega_{\Lambda},
\\
\lb{8b}
\vec{P}_{\Lambda}\Omega_{\Lambda} &=& \vec{0}.
\eeqa
\end{subequations}

\noindent The fact that the ground state is simple, i.e.,
nondegenerate, is a property of Bosons only which will be used later
(see, e.g., \cite{martin} for a popular textbook account: for precise
conditions on $\Phi$ in (\ref{4}) see \cite{reed}, where a proof is
given). By (\ref{6}) - (\ref{8}), 

\begin{subequations}
\lb{9}
\beq
H_{\Lambda}^{\vec{v}}\Omega_{\Lambda} = 
\Bigg( E_{0}(N,L)+\frac{1}{2}N\vec{v}^2 \Bigg) \Omega_{\Lambda}. \lb{9a}
\eeq

\noindent Thus the energy eigenvalue of $\Omega_\Lambda$ in frame $F$ is

\beq
E_{0}^{F}\equiv E_{0}(N,L)+\frac{1}{2}N\vec{v}^2. \lb{9b}
\eeq
\end{subequations}

\noindent Because of (\ref{6}), there exists common eigenstates of
$H_{\Lambda}$ and $\vec{P}_{\Lambda}$, which we denote by
$\psi_{\Lambda}^{\vec{k}}$, i.e., 

\beqa
\lb{10}
H_{\Lambda}\psi_{\Lambda}^{\vec{k}} &=& 
E_{\Lambda}(\vec{k})\psi_{\Lambda}^{\vec{k}}, 
\\
\lb{11}
\vec{P}_{\Lambda}\psi_{\Lambda}^{\vec{k}} &=&
\vec{k}\psi_{\Lambda}^{\vec{k}},
\eeqa

\noindent where $\vec{k}\in\Lambda^{\ast}$, and 

\beq
\lb{12}
\Lambda^{\ast}\equiv 
\Bigg\{\frac{2\pi}{L}\vec{n}, \vec{n}\equiv(n_1,n_2,n_3);
n_i\in{\mathbb Z}, i=1,2,3 \Bigg\}.
\eeq 

\noindent Let $\psi_{\Lambda}^{(0),\vec{k}}$ be common eigenstates of
$H_{\Lambda}$ and $\vec{P}_{\Lambda}$ of the smallest energy above
$E_{0}(N,L)$, which we denote by $E_{0}^{\Lambda}(\vec{k})$. Under
certain conditions, seen to be realized in soluble models
(\cite{girardeau}, \cite{lieb1}, \cite{lieb2}), it may be expected
that 

\beq
\lb{13}
E_{0}^{\Lambda}(\vec{k}) = E_{0}(N,L) + \epsilon_{\Lambda}(\vec{k}),
\eeq

\noindent where $\epsilon_{\Lambda}(\vec{k})$ is the energy of an
``elementary excitation'' (see \cite{lieb2} for a better definition of
this important concept, which should not be confused with
``quasiparticle'' - the latter dissipate). By (\ref{7}) and
(\ref{13}), the energy of the lowest energy excitations above the
ground state in frame $F$ is given by

\beq
\lb{14}
E_{0}(N,L) + \epsilon_{\Lambda}(\vec{k}) +\vec{v}\cdot\vec{k} +
\frac{1}{2}N\vec{v}^2.
\eeq 

\noindent There will be \underline{dissipation} if this energy is
smaller than the energy of $\Omega_\Lambda$ in frame $F$, which is
given by (\ref{9b}); thus, by (\ref{9b}) and (\ref{14}), there is no 
dissipation, i.e., there is superfluidity, if

\beq
\lb{15}
\epsilon_{\Lambda}(\vec{k}) + \vec{v}\cdot\vec{k} \geq 0,
\eeq

\noindent which is expected to hold if (\ref{1}) is satisfied. This is
\underline{Landau's criterion}. It is satisfied (with a calculable
$v_c$) in the two remarkable models (\cite{girardeau}, \cite{lieb1},
\cite{lieb2}). The above relation (\ref{15}) also provides the
condition for $\Omega_\Lambda$ to be the ground state in the moving
frame. 

In order to formulate Landau's criterion more precisely, we have to
define the thermodynamic limit of $\epsilon_{\Lambda}(\vec{k})$ for
any $\vec{k}$, not only those in (\ref{12}). Let $\{\Lambda_n\}$ be an
increasing sequence of periodic boxes, 

\begin{subequations}
\lb{16}
\beq
\lb{16a}
\vec{k}_n\in\Lambda^{*}_n \quad\text{and}\quad
\vec{k}_n\longrightarrow\vec{k}\neq\vec{0}.
\eeq 

\noindent Since any real number can be expressed as a limit of
rational numbers, we may define: 

\beq
\lb{16b}
\widetilde{\epsilon}(\vec{k})=\lim_{n}\epsilon_{\Lambda_n}(\vec{k}_n)
\eeq
\end{subequations}

\noindent on the above defined $\widetilde{\epsilon}(\vec{k})$ we
formulate then:

\vspace{.5cm}

\noindent \underline{{\bf Assumption i}} -
$\widetilde{\epsilon}(\vec{k})>0$ for $\vec{k}\neq\vec{0}$ is
independent of the sequence in (\ref{16b}) and is a continuous
function of $\vec{k}$ when the limit on the r.h.s. of (\ref{16b})
exists. 

\vspace{.5cm}

By (\ref{15}), (\ref{16}) and assumption {\bf i}, the following
inequality holds for all $\vec{k}\in{\mathbb R}^3$:

\beq
\lb{17}
\widetilde{\epsilon}(\vec{k}) + \vec{v}\cdot\vec{k}\geq 0.
\eeq

\noindent The above may look too pedantic, but we shall see that the
thermodynamic limit  will play an important role later on. With the
above assumptions, we shall drop the tilde on $\epsilon(\vec{k})$ from
now on. We remark that assumption {\bf i} has been proved
in at least one case, that of magnons (or spin-waves) in quantum
ferromagnets (\cite{vanhemmen}, \cite{streater}).

An important attempt to ``explain'' (\ref{17}), i.e., to derive
(\ref{17}) from microscopic laws, was undertaken by Bogoliubov (see
\cite{zagrebnov} for an excellent review), assuming the existence of
Bose Einstein Condensation (BEC). As remarked by Leggett in his
beautiful review \cite{leggett}, ``while extremely suggestive,
Bogoliubov's result referred to a dilute system, which is rather far
from real-life liquid {\bf He-II}''. What seems to be usually not
emphasized is that Bogoliubov's model, even in the version which
conserves particle number \cite{zagrebnov}, is \underline{not}
Galilean-covariant, i.e., does not satisfy (\ref{7}), and, thus, does
not satisfy local mass conservation, an essential physical requirement
(this seems to be well known but, for a proof, see \cite{wreszinski}).
Bogoliubov's seminal work led, however, to the genesis of an
understanding of $e_{0}(\rho)$ (defined by (\ref{26})) through later
work (see \cite{lieb3} and references given there). One of its
important byproducts was the realization that it is the
\underline{repulsive} part of the potential which plays the major role
in superfluidity. This is vindicated in the remarkable one-dimensional
models of Girardeau \cite{girardeau} and Lieb and Liniger
(\cite{lieb1}, \cite{lieb2}). The latter, however, illustrate an
important fact: superfluidity is \underline{independent} of BEC,
because these models display superfluidity in the sense of Landau
\cite{lieb2}, but no BEC (this was proved in \cite{lenard} for
Girardeau's model, and is presumably also true for the Lieb-Liniger
model).

As remarked by Leggett in \cite{leggett}, a particular successful
attack on the full {\bf He-II} problem (i.e., not only the dilute
case) was made by Feynman \cite{feynman1} and Feynman-Cohen
\cite{feynman2} (see also Lieb's remarkable early review \cite{lieb3})
through a variational wave-function (in our notation (\ref{8}),
(\ref{10}), (\ref{11})): 

\beq
\lb{18}
\psi_{\Lambda}^{\vec{k}}(\vec{x}_1,\ldots,\vec{x}_N)\equiv\sum_{i=1}^{N}
\mb{e}^{\text{i}\vec{k}\cdot\vec{x}_i}\Omega_{\Lambda}(\vec{x}_1,\ldots,
\vec{x}_N),
\eeq

\noindent (the Feynman-Cohen Ansatz \cite{feynman2} leads to somewhat
better results, but we shall not consider it in this paper). By using
partial integration, and the periodic b.c. (see, e.g., \cite{martin},
Exercise p. 262, or \cite{feynman1}), we obtain

\beq
\lb{19}
{\la\psi_{\Lambda}^{\vec{k}}|H_{\Lambda}\psi_{\Lambda}^{\vec{k}}\ra\over
\la\psi_{\Lambda}^{\vec{k}}|\psi_{\Lambda}^{\vec{k}}\ra} = E_{0}(N,L) + 
\mathcal{E}_{\Lambda}(\vec{k}),
\eeq

\noindent where

\beq
\lb{20}
\mathcal{E}_{\Lambda}(\vec{k})\equiv\frac{N\vec{k}^2}
{2\la\psi_{\Lambda}^{\vec{k}}|\psi_{\Lambda}^{\vec{k}}\ra} =
\frac{\vec{k}^2}{2\mathcal{S}_{\Lambda}(\vec{k})}
\eeq

\noindent and

\beq
\lb{21}
\mathcal{S}_{\Lambda}(\vec{k}) = \frac{1}{N}
\la\psi_{\Lambda}^{\vec{k}}|\psi_{\Lambda}^{\vec{k}}\ra 
= \frac{1}{N}\int_{\Lambda^{N}}d\vec{x}_1\cdots d\vec{x}_N
\Bigg|\sum_{i=1}^{N}\mb{e}^{-\text{i}\vec{k}\cdot\vec{x}_i}\Bigg|^2 
\Omega_{\Lambda}^{2}(\vec{x}_1,\ldots,\vec{x}_N),  
\eeq

\noindent (As follows from the Perron-Frobenius theorem,
$\Omega_{\Lambda}$ is unique, and may be taken to be positive
\cite{reed}). Except for a factor $\rho=\frac{N}{V}$, (\ref{21})
defines the \underline{liquid structure factor}, which may - in case
of liquid {\bf He-II} - be measured independently of the excitation
spectrum, by X-ray scattering (see \cite{huang}, Fig. 13.10). 

By (\ref{21}), $\mathcal{S}_{\Lambda}(\vec{k})$ has a direct physical
interpretation: it is - for $\vec{k}\neq\vec{0}$ -  the Fourier
transform of the pair correlation function in the ground state
(\cite{fisher1}, pg. 951).

Equation (\ref{18}) and the min-max principle imply that, for
$\vec{k}\neq\vec{0}$, (\ref{19}) is an \underline{upper} bound to the
energies of the ``elementary excitations'', i.e., the lowest
eigenvalues of $H_{\Lambda}$ which are larger than the ground state
energy. The following assumption will be made:

\vspace{.5cm}

\noindent \underline{{\bf Assumption ii}} - The Feynman Ansatz
(\ref{18}) yields a (qualitatively) good approximation to the energy
of the elementary excitations.

\vspace{.5cm}

\noindent The above assumption is widely accepted today
\cite{leggett}, because it is confirmed by experiment both for liquid
{\bf He-II} \cite{huang} and for trapped gases \cite{steinhauer2}.

\vspace{.5cm}

\noindent Assumptions {\bf i} and {\bf ii} imply, together with (\ref{20}) that  

\vspace{.3cm}

\noindent \underline{{\bf Assumption i'}} - Under (\ref{16a}), the limit 

\beq
\lb{22}
\mathcal{S}(\vec{k})\equiv\lim_{n}\mathcal{S}_{\Lambda_n}(\vec{k}_n),
\eeq

\noindent exists and is a continuous function of $\vec{k}$. 

\vspace{.5cm}

\noindent The fact that it is necessary to take the thermodynamic
limit with $\vec{k}\neq\vec{0}$ is shown in Appendix A for the free
Bose gas: the limits $\vec{k}\rightarrow\vec{0}$ and the thermodynamic
limit do not commute. This procedure has also a physical reason: it is
necessary to eliminate the (infinite) forward scattering peak, i.e.,
by considering $\vec{k}\neq\vec{0}$, before probing the
density-density correlations for the infinite system (\cite{fisher1},
pg. 951). 

\renewcommand{\thesubsection}{\thesection.\arabic{subsection}}

\subsection{Onsager's Inequality}

The following inequality (\ref{23}) (with condition (\ref{24})) will
be referred to as Onsager's inequality \cite{price}:

\beq
\lb{23}
\mathcal{S}(\vec{k})\leq c|\vec{k}|;\quad\quad \vec{k}\neq\vec{0},
\eeq

\noindent where

\beq
\lb{24}
0<c<\infty
\eeq

\noindent is a constant independent of $N$, $L$, but possibly
dependent on $\rho$. 

We see from the above definition that Onsager's inequality, taken
together with Feynman's Ansatz (Assumption {\bf ii} and (\ref{18}))
implies Landau's condition (\ref{17}), with $v_c$ (defined by (1))
given by

\beq
\lb{25}
v_c = (2c)^{-1}.
\eeq

\noindent For the free Bose gas $S_\Lambda(\vec{k})$, given by
(\ref{21}), equals 1 (see (\ref{a2}) of Appendix A), which is, of
course, necessary in order to be compatible with (\ref{20}).

Let $e_0(\rho)$ denote the ground state energy per particle in the
thermodynamic limit

\beq
\lb{26}
e_0(\rho) = \lim_{L\rightarrow\infty}{E_0(\rho L^3,L)\over(\rho L^3)}.
\eeq
 
\noindent By \cite{fisher2} $e_0(\rho)$ is a convex function of
$\rho$. Thus, when ${e_0}''(\rho)\equiv d^2e_0(\rho)/d\rho^2$ exists, 

\beq
\nnb
{e_0}''(\rho)\geq 0.
\eeq

\noindent We assume that

\beq
\lb{27}
0<{e_0}''(\rho)<\infty.
\eeq

\noindent (\ref{23}) and (\ref{24}) constitute a precise formulation
of \underline{Onsager's inequality} \cite{price}, with

\beq
\lb{28}
c\equiv\Bigg({1\over\rho{e_0}''(\rho)}\Bigg)^{1/2}.
\eeq

\noindent Note that (\ref{24}) and (\ref{28}) are consistent due to
(\ref{27}) if we assume $0<\rho<\infty$ which we do henceforth.

In this paper we derive (\ref{23}), (\ref{24}) for a soluble model,
Girardeau's model \cite{girardeau}, defined in section {\bf IV},
which is believed to be a prototype of a class of one-dimensional Bose
fluids with pointwise repulsive interactions. We believe that several
aspects of the derivation are relevant to higher dimensions and
discuss this in section {\bf V}. In particular, Proposition 1 of
section {\bf III} is of general validity, as shown in
\cite{price}. Our derivation in section {\bf III} ``rounds off'' some
points left over in \cite{price}, such as emphasizing the important
role of nondegeneracy of the ground state - which leads to
nondegenerate perturbation theory - and the vanishing of some mixed
terms'' (\ref{36}). In section {\bf IV} we conclude the proof of
(\ref{23}), (\ref{24}) for Girardeau's model. 

It is important to realize the intrinsically nonperturbative nature
(in $\vec{k}$) of the bound (\ref{23}) on
$\mathcal{S}_{\Lambda}(\vec{k})$ in the thermodynamic limit. Writing
in (\ref{21}),

\beq
\lb{29}
\Bigg|\sum_{i=1}^{N}\mb{e}^{-\text{i}\vec{k}\cdot\vec{x}_i}\Bigg|^2
= 1 + \sum_{\overset{i,j=1}{i\neq j}}^{N} 
\mb{e}^{-\text{i}\vec{k}\cdot(\vec{x}_i-\vec{x}_j)}, 
\eeq

\noindent it becomes clear that the double sum above would contribute
a term $O(N^2)/N=O(N)$ to $\mathcal{S}_{\Lambda}(\vec{k})$ if the
configurations $\vec{x}_i\approx\vec{x}_j$, $\forall i\neq j$, were not
suppressed: due to the strong repulsion at short distances, such
configurations are highly improbable in
$\Omega(\vec{x}_1,\ldots,\vec{x}_N)$ (think of hard cores). But the
linear term in the \underline{perturbative} expansion (in $\vec{k}$)
of (\ref{29}) is identically zero by symmetry, thus suggesting
$\mathcal{S}_{\Lambda}(\vec{k})=O(|\vec{k}|^2)$!

\section{Proof of an Auxiliary Proposition}

We now provide a proof of an auxiliary proposition (Proposition
1). The derivation in \cite{price} is incomplete in some details. Let
$P_{0,\Lambda}$ denote the projector onto the orthogonal complement
$\Omega_{\Lambda}^{\perp}$ of the unique ground state
$\Omega_{\Lambda}$, $\psi_{\Lambda}^{(\vec{k})}$ be given by
(\ref{18}), and 

\beqa
\lb{30}
A_{\Lambda}&\equiv& P_{0,\Lambda}(H_{\Lambda} -
E_0(N,L))^{1/2}P_{0,\Lambda}, \\
\lb{31}
B_{\Lambda}&\equiv& P_{0,\Lambda}(H_{\Lambda} -
E_0(N,L))^{-1/2}P_{0,\Lambda}.
\eeqa

\noindent Since, by (\ref{8}) and (\ref{18}),
$\psi_{\Lambda}^{\vec{k}}$ is an eigenfunction of $\vec{P}_{\Lambda}$
with eigenvalue $\vec{k}$, and $\vec{k}\neq\vec{0}$,
$\psi_{\Lambda}^{\vec{k}}\in
P_{0,\Lambda}\mathcal{H}_{\Lambda}$, and thus, by the
Schwarz inequality

\beq
\lb{32}
\la\psi_{\Lambda}^{\vec{k}}|\psi_{\Lambda}^{\vec{k}}\ra^2 =
\la
A_{\Lambda}\psi_{\Lambda}^{\vec{k}}|B_{\Lambda}\psi_{\Lambda}^{\vec{k}}\ra^2
\leq
\la A_{\Lambda}\psi_{\Lambda}^{\vec{k}}|A_{\Lambda}\psi_{\Lambda}^{\vec{k}}\ra
\la B_{\Lambda}\psi_{\Lambda}^{\vec{k}}|B_{\Lambda}\psi_{\Lambda}^{\vec{k}}\ra
\eeq

\noindent but

\beqa
\nnb
\la A_{\Lambda}\psi_{\Lambda}^{\vec{k}}|A_{\Lambda}\psi_{\Lambda}^{\vec{k}}\ra
&=&
\la P_{0,\Lambda}\psi_{\Lambda}^{\vec{k}}|(H_{\Lambda} - E_0(N,L))
P_{0,\Lambda}\psi_{\Lambda}^{\vec{k}}\ra \\
\nnb
&=&
\la\psi_{\Lambda}^{\vec{k}}|(H_{\Lambda} -
E_0(N,L))\psi_{\Lambda}^{\vec{k}}\ra \\
\lb{33}
&=& 
N\frac{\vec{k}^2}{2}
\eeqa

\noindent by partial integration and use of the periodic b.c.. We also
have that

\begin{widetext}
\beqa
\nnb
\la B_{\Lambda}\psi_{\Lambda}^{\vec{k}}|B_{\Lambda}\psi_{\Lambda}^{\vec{k}}\ra
&=&
\la\psi_{\Lambda}^{\vec{k}}|(H_{\Lambda} -
E_0(N,L))^{-1/2}P_{0,\Lambda}(H_{\Lambda} -
E_0(N,L))^{-1/2}\psi_{\Lambda}^{\vec{k}}\ra \\
\lb{34}
&=&
\la\Omega_{\Lambda}|{W_{N}^{\vec{k}}}^{\ast}(H_{\Lambda} -
E_0(N,L))^{-1}P_{0,\Lambda}W_{N}^{\vec{k}}\Omega_{\Lambda}\ra.
\eeqa
\end{widetext}

\noindent In (\ref{34}), $W_{N}^{\vec{k}}$ is the multiplication
operator

\beq
\lb{35}
(W_{N}^{\vec{k}}\varphi)(\vec{x}_1,\ldots,\vec{x}_N) = \sum_{i=1}^{N}
\mb{e}^{\text{i}\vec{k}\cdot\vec{x}_i}\varphi(\vec{x}_1,\ldots,\vec{x}_N).
\eeq

\noindent We now write
$W_{N}^{\vec{k}}=\sum_{i=1}^{N}\cos(\vec{k}\cdot\vec{x}_i)
+\text{i}\sum_{i=1}^{N}\sin(\vec{k}\cdot\vec{x}_i)$ in (\ref{34}). For
finite $N$, $L$, $(H_{\Lambda}-E_0(N,L))$ has a nonzero lower bound on
$P_{0,\Lambda}\mathcal{H}_{\Lambda}$, and insertion of a basis of
eigenstates $|\phi_i\ra$ of $(H_{\Lambda}-E_0(N,L))$ in
$P_{0,\Lambda}\mathcal{H}_{\Lambda}$ is rigorously justified because
$\Omega_{\Lambda}$ is a normalized state. Furthermore, because
$H_{\Lambda}$ is a real operator, we may choose this basis of
eigenstates $|\phi_i\ra$ as consisting of \underline{real}
functions. The ``mixed terms'' in (\ref{34})

\begin{subequations}
\lb{36}
\beq
\lb{36a}
\sum_{j}\frac{\text{-i}}{E_j-E_0}\la\Omega_{\Lambda}|
\sum_{i=1}^{N}\sin(\vec{k}\cdot\vec{x}_i)|\phi_j\ra\la\phi_j|
\sum_{i=1}^{N}\cos(\vec{k}\cdot\vec{x}_i)|\Omega_{\Lambda}\ra 
\eeq

\noindent and

\beq
\lb{36b}
\sum_{j}\frac{\text{i}}{E_j-E_0}\la\phi_j|
\sum_{i=1}^{N}\cos(\vec{k}\cdot\vec{x}_i)|
\Omega_{\Lambda}\ra\la\Omega_{\Lambda}|
\sum_{i=1}^{N}\sin(\vec{k}\cdot\vec{x}_i)|\phi_j\ra
\eeq
\end{subequations}

\noindent thus add to zero, and we obtain from (\ref{34}) and analytic
perturbation theory \cite{kato}:

\noindent\underline{{\bf Proposition 1}}\,:

\begin{subequations}
\lb{37}
\beq
\lb{37a}
\la\psi^{\vec{k}}_{\Lambda}|\psi^{\vec{k}}_{\Lambda}\ra^2
\leq\frac{N\vec{k}^2}{2}\la B_{\Lambda}\psi_{\Lambda}^{\vec{k}}|
B_{\Lambda}\psi_{\Lambda}^{\vec{k}}\ra
\eeq

\noindent where

\beq
\lb{37b}
\la B_{\Lambda}\psi_{\Lambda}^{\vec{k}}|
B_{\Lambda}\psi_{\Lambda}^{\vec{k}}\ra = 
\frac{1}{2}\Bigg(-\frac{\partial^2}{\partial\lambda^2}
E_{0,1}(N,L,\lambda)\Bigg)_{\lambda =0} + 
\frac{1}{2}\Bigg(-\frac{\partial^2}{\partial\lambda^2}E_{0,2}
(N,L,\lambda)\Bigg)_{\lambda =0} 
\eeq
\end{subequations}

\noindent Above, $E_{0,1}(N,L,\lambda)$ is the ground state energy of the
Hamiltonian

\begin{subequations}
\lb{38}
\beq
H_{\Lambda}^{(1)}(\lambda)\equiv
H_{\Lambda}-E_0(N,L)+\lambda\sum_{i=1}^{N}\cos(\vec{k}\cdot\vec{x}_i)
\lb{38a}
\eeq

\noindent and $E_{0,2}(N,L,\lambda)$ is the ground state energy of the
Hamiltonian

\beq
H_{\Lambda}^{(2)}(\lambda)\equiv
H_{\Lambda}-E_0(N,L)+\lambda\sum_{i=1}^{N}\sin(\vec{k}\cdot\vec{x}_i).
\lb{38b}
\eeq
\end{subequations}

\noindent By (\ref{34}) and the argument following it the two
quantities on the r.h.s. of (\ref{37}) are also the second order
energy terms in analytic perturbation theory
(\cite{kato}, \cite{galindo}). The boson nature of the particles is used to
the \underline{sole extent} that the ground state is nondegerate
\cite{reed}. The radius of convergence of the perturbation series in
$\lambda$ is expected to tend to zero as $(1/N)$, for our $N$-particle
system, but this is no source of trouble as long as $\lambda$ is taken
to go to zero in (\ref{37}) for fixed, finite $N$ and $L$. What is
remarkable in (\ref{37}), (\ref{38}) is the fact that the perturbation
is an \underline{external potential}. 

\section{Onsager's Inequality and Girardeau's Model}

A heuristic but appealing argument due to Onsager (see \cite{price}
and appendix B) yields

\beq
\lb{39}
\lim_{\overset{N\rightarrow\infty}{\overset{L\rightarrow\infty}
{\frac{N}{L^3}=\rho}}}
\Bigg[\frac{1}{N}
\la B_{\Lambda}\psi_{\Lambda}^{\vec{k}}|B_{\Lambda}\psi_{\Lambda}^{\vec{k}}\ra
\Bigg]\leq c_1
\eeq

\noindent where 

\beq
\lb{40}
c_1\equiv\frac{1}{\rho e_0''(\rho)}
\eeq

\noindent whenever the limit on the l.h.s. of (\ref{39})
exists. Putting together (\ref{21}), (\ref{32}), (\ref{33}),
(\ref{39}) and (\ref{40}), we arrive at (\ref{25}), (\ref{28}) with
$c=(c_1/2)^{1/2}$. 

Other arguments (see, e.g., (A.97)-(A.103) of \cite{huang},
``longitudinal sum rules'') which yield $\mathcal{S}(\vec{k})\sim
c|\vec{k}|$ for $|\vec{k}|$ sufficiently small, all depend on the
argument in Appendix B. This is a density-functional type of
argument, which, in general, is not generally justifiable for
Fermions, whenever the ground state is degenerate \cite{lieb4}. For
Bosons, the ground state is nondegenerate, and density-functional theory
looks the same as for Fermions, but it has been rigorously established
recently that, even in the dilute limit, a semiclassical description
of Bosons is impossible (see, e.g., \cite{lieb5}, remarks after
(2.12)). Thus, the derivation of (\ref{39}), (\ref{40}) given in
\cite{price} (a somewhat better version of which, with corrections, is
given in appendix B) is conceptually open to question. But, perhaps
most importantly, the derivation has a ``miraculous'' character,
because in (\ref{34}) the contribution of intermediate states is
expected to yield $(const.)\times|\vec{k}|^{-1}$ due to the
denominator $(H_{\Lambda}-E_0(N,L))^{-1}$ and the fact that the lowest
states are expected to have energy $\epsilon_{\Lambda}(\vec{k})\sim
c|\vec{k}|$ above the ground state! These matters are dealt in a
different way in the appendix of \cite{feynman2}: there, the argument
relies on the study of the terms of a perturbation theory which likely
diverges. 

We now examine (\ref{39}), (\ref{40}) on the light of one of the very
few soluble models of a superfluid, the Girardeau model (the
Lieb-Liniger model (\cite{lieb1}, \cite{lieb2}) is expected to yield
similar results). We shall see that the results confirm (\ref{39}),
(\ref{40}) surprisingly well, and formula (\ref{b11}), derived from
density functional theory, holds exactly, in the limit
$\vec{k}\rightarrow\vec{0}$.     

We start by the Hamiltonian of the Lieb-Liniger model \cite{lieb1}

\beq
\lb{41}
H_{N,L}^{d} = -\frac{1}{2}\sum_{i=1}^{N}
\frac{\partial^2}{\partial x_i^2} +
2d\sum_{1\leq i<j\leq N}\delta(x_i - x_j)
\eeq

\noindent with $d>0$, with periodic b.c. on a box of length $L$. In the
limit $d\rightarrow\infty$ one obtains a model of impenetrable bosons,
with the impenetrable core shrunk to a point, previously treated by
Girardeau \cite{girardeau}, and called Girardeau's model. Both models
exhibit two branches of Bogoliubov excitations which satisfy Landau's
criterion \cite{lieb2}, and which recently have been seen
experimentally \cite{steinhauer1}. Since the results on Girardeau's
model may be obtained from the Lieb-Liniger model (\ref{41}) by
continuity as $d\rightarrow\infty$ (\cite{lieb1}, \cite{lieb2}), we
expect that our results are also true for the richer model (\ref{41}),
and are, thus, prototypical of this class of models.

One of the basic features of Girardeau's model is that, due to the
repulsive interaction, the system acquires a \underline{finite}
compressibility (proportional to $(e_0''(\rho))^{-1}$), in contrast to
the free Bose gas, for which, e.g., with Dirichlet b.c., 

\beq
\nnb
E_0(\rho L^3,L)=f\rho L
\eeq

\noindent where $f$ is a constant, and thus, by (\ref{28}),
$e_0(\rho)=0$ for all $\rho$. Thus $e_0''(\rho)$ is also identically
zero, i.e., $c_1=+\infty$ in (\ref{39}), (\ref{40}). Thus the
inequality (\ref{23}) (with $c$ given by (\ref{28})) is trivially true
by (\ref{a2}). 

Performing the limit $d\rightarrow\infty$ on (\ref{41}) is equivalent
to impose the subsidiary condition:

\beq
\lb{42}
\psi(x_1,\ldots,x_N)=0\quad\text{if}\quad x_j=x_l,\quad 1\leq j<l\leq N.
\eeq

\noindent The Bose eigenfunctions $\psi^{B}$ satisfying (\ref{42}) and
periodic b.c. with period $L$ are given \cite{girardeau} as

\begin{subequations}
\lb{43}
\beq
\lb{43a}
\psi^B=\psi^F\mathcal{A}
\eeq

\noindent where

\beq
\lb{43b}
\mathcal{A}(x_1,\ldots,x_N)=\prod_{j>l}\mb{sgn}(x_j-x_l).
\eeq
\end{subequations}

\noindent Above, $\mb{sgn}(x)$ is the algebraic sign of $x$, and
$\psi^F$ are Fermi energy eigenfunctions satisfying (\ref{42}), which
are just the eigenfunctions of a \underline{free} Fermi gas. For odd N
we fill the ``Fermi Sphere''

\beq
\lb{44}
-\frac{1}{2}(N-1)\leq p\leq\frac{1}{2}(N-1)
\eeq

\noindent and the ground state energy is 

\beq
\lb{45}
E_0(N,L)=\sum_{p=1}^{\frac{1}{2}(N-1)}(\frac{2\pi p}{L})^2 =
\frac{1}{6}(N-N^{-1})(\pi\rho)^2.
\eeq

\noindent Thus the ground state energy density in the thermodynamic
limit equals 

\beq
\lb{46}
e_0(\rho)=\frac{1}{6}\pi^2\rho^3
\eeq

\noindent and thus

\beq
\lb{47}
e_0''(\rho)=\pi^2\rho
\eeq

\noindent which satisfies (\ref{27}).

\noindent By (\ref{37}), (\ref{38}), (\ref{44}) 
$\la B_{\Lambda}\psi_{\Lambda}^{k}|
B_{\Lambda}\psi_{\Lambda}^{k}\ra$ may be calculated by the
formula ($N$ odd)

\beq
\lb{48}
\la B_{\Lambda}\psi_{\Lambda}^{k}|B_{\Lambda}\psi_{\Lambda}^{k}\ra
=
\frac{1}{2}\Bigg\{\Bigg(-\frac{\partial^2}{\partial\lambda^2}E_{0,1}
(N,L,\lambda)\Bigg)_{\lambda=0} + 
\Bigg(-\frac{\partial^2}{\partial\lambda^2}E_{0,2}(N,L,\lambda)\Bigg)_{\lambda
=0}\Bigg\}
\eeq

\beq
\lb{49}
E_{0,1}(N,L,\lambda) =
\sum_{p=-\frac{1}{2}(N-1)}^{\frac{1}{2}(N-1)}E_1(p,\lambda) 
\eeq

\beq
\lb{50}
E_{0,2}(N,L,\lambda) =
\sum_{p=-\frac{1}{2}(N-1)}^{\frac{1}{2}(N-1)}E_2(p,\lambda)
\eeq

\noindent where $E_1(p,\lambda)$ are the energy levels of the
one-particle Hamiltonian

\beq
\lb{51}
H=H_0 +\lambda H_1
\eeq

\noindent with

\beq
\lb{52}
H_0 = -\frac{1}{2}\frac{d^2}{dx^2}
\eeq

\noindent and 

\beq
\lb{53}
H_1 = \cos(kx)
\eeq

\noindent on $\mathcal{H}=L_{\text{per}}^2 [0,L]$, with
$k\in\Lambda^{\ast}=\{\frac{2\pi}{L}n,\quad n\in{\mathbb Z}\setminus\{
0\}\}$ and the zero$^{\underline{th}}$ order level of $H_0$ corresponding to
$E_1(p,\lambda)$ is

\beq
\lb{54}
E_1^{(0)}(p) = \frac{1}{2}(\frac{2\pi p}{L})^2
\eeq

\noindent with $p$ satisfying (\ref{44}). The levels $E_2(p,\lambda)$
are the same, just replacing $H_1$ by $H_2=\sin(kx)$. The level $p=0$
of $H_0$ is nondegenerate, all the others in (\ref{44}) are doubly
degenerate, with energies (\ref{54}) and eigenfunctions

\beq
\lb{55}
\varphi_{p,\alpha=1}=\frac{1}{\sqrt{L}}\mb{e}^{\mb{i}\frac{2\pi
p}{L}x}\,,\quad
\varphi_{p,\alpha=2}=
\frac{1}{\sqrt{L}}\mb{e}^{-\mb{i}\frac{2\pi p}{L}x}\,;\quad\quad 
p\in\{1,2,\ldots,\frac{N-1}{2}\}.
\eeq

\noindent By Kato degenerate perturbation theory (\cite{kato},
\cite{galindo} (10.88)) we find the levels up to second order by
solving the equation

\beq
\lb{56}
\det\{\lambda\la\varphi_{p,\beta}|H_1|\varphi_{p,\alpha}\ra + 
\lambda^2\la\varphi_{p,\beta}|H_1 S_p H_1|\varphi_{p,\alpha}\ra - 
(E_1(p,\lambda)-E_1^{(0)})\delta_{\beta,\alpha}\}=0 
\eeq

\noindent where

\beq
\lb{57}
S_p\equiv\frac{Q_p^{(0)}}{E_1^{(0)}{\mathbb I}-H_0},\quad\quad Q_p^{(0)}=
{\mathbb I}-P_p^{(0)}
\eeq 

\noindent and 

\beq
\lb{58}
P_p^{(0)} = |\varphi_{p,1}\ra\la\varphi_{p,1}| +
|\varphi_{p,2}\ra\la\varphi_{p,2}|. 
\eeq

Using (\ref{55}), (\ref{57}), (\ref{58}) and plane-waves as
intermediate states to calculate the second term in (\ref{56}), we
find that equation (\ref{56}) may be written

\beq
\lb{59}
\det\begin{pmatrix}a-\mu & c \\ c & a-\mu\end{pmatrix} =0 
\eeq

\noindent with

\beqa
\lb{60}
&&\mu\equiv E_1(p,\lambda)-E_1^{(0)} \\
\nnb \\
\lb{61}
&&a = \frac{-\lambda^2}{k^2-4(\frac{2\pi}{L}p)^2}
\delta_{\frac{2\pi}{L}p\neq\frac{k}{2}} \\
\nnb \\
\lb{62}
&&c = \frac{\lambda}{2}\delta_{k,\frac{4\pi}{L}p} +
\frac{\lambda^2}{4\varepsilon_k}\delta_{\frac{2\pi}{L}p,k}\quad;
\quad\varepsilon_k=\frac{k^2}{2}.
\eeqa   

\noindent The discriminant of (\ref{59}) is $4a^2-4(a^2-c^2)=4c^2 >0$,
and thus the eigenvalues for $\mu$ are

\beq
\lb{63}
\mu_{\pm}=a\pm c.
\eeq

\noindent The cases $p=0$, $\frac{2\pi p}{L}=k$, and $\frac{2\pi
p}{L}=\frac{k}{2}$ are isolated values which by (\ref{48}), (\ref{49})
do not affect the thermodynamic limit on the l.h.s. of
(\ref{39}). Disregarding these, we see by (\ref{63}) that the
eigenvalues remain doubly degenerate and equal to $a$, given by
(\ref{61}). Thus, by (\ref{48}), (\ref{49}) and (\ref{61}), and the
fact that the second term in (\ref{48}) is equal to first, which may
be proved, we find 

\beq
\lb{64}
e_{\Lambda}(k)\equiv\frac{1}{N}
\la B_{\Lambda}\psi_{\Lambda}^{k}|B_{\Lambda}\psi_{\Lambda}^{k}\ra =
4(\frac{L}{2\pi N})(\frac{2\pi}{L})
\sum_{\overset{p=-\frac{1}{2}(N-1)}{\overset{}{\overset{p\neq0}
{\overset{}{\frac{2\pi p}{L}\neq\pm\frac{k}{2}}}}}}^{\frac{1}{2}(N-1)}
\frac{1}{k^2-4(\frac{2\pi p}{L})^2}.  
\eeq

\noindent Clearly the r.h.s. of (\ref{64}) has a natural extension
$\widetilde{e}_{\Lambda}(k)$ to all $k\neq0$, obtained by interpreting
$k$ in (\ref{64}) as a real variable. We now find, from (\ref{64}) the
following proposition proved in Appendix C:  

\noindent\underline{{\bf Proposition 2}}\,:

\beq
\lb{65}
\widetilde{e}(k)\equiv\lim_{\overset{N\rightarrow\infty}{\overset{L\rightarrow\infty}
{\frac{N}{L}=\rho}}}\widetilde{e}_{\Lambda}(k) = \frac{4}{2\pi\rho}
Pv\int_{-\pi\rho}^{\pi\rho}\frac{1}{k^2-4x^2}dx. 
\eeq

\noindent  {\it Remark 1}. The principal value on the r.h.s. of (\ref{65}) may
computed, with the result 

\beq
\lb{66}
Pv\int_{-\pi\rho}^{\pi\rho}\frac{1}{k^2-4x^2}dx =
\frac{1}{|k|}\mb{arccoth}\Bigg(\frac{2\pi\rho}{|k|}\Bigg).
\eeq

\noindent The limit $k\rightarrow0$ on the r.h.s. of (\ref{66}) can be
performed by l'H\^opital's rule or more simply by noticing that
$\text{arccoth}(2\pi\rho/|k|)\approx |k|/2\pi\rho$, so that

\beq
\lb{67}
\lim_{k\rightarrow0}\frac{1}{|k|}\mb{arccoth}\Bigg(\frac{2\pi\rho}{|k|}\Bigg)
= \frac{1}{2\pi\rho}.
\eeq

\noindent We finally have

\noindent {\bf Theorem 1}. For Girardeau's model, under assumption
{\bf i'} $\mathcal{S}(k)$ satisfies Onsager's inequality (\ref{23}),
(\ref{24}), with 

\begin{subequations}
\lb{68}
\beq
\lb{68a}
c=\frac{1}{\sqrt{2}}\frac{1}{\pi\rho}t
\eeq

\noindent where

\beq
\lb{68b}
t\simeq 1.0481.
\eeq
\end{subequations}

\noindent {\it Proof}. By Proposition 2, for any sequence
$\{\Lambda_n\}$ of periodic boxes, the limit
$\lim_{n}e_{\Lambda_n}(k)$ is a continuous function of
$k\in[0,\pi\rho]$ and it follows from appendix C that the convergence
$\widetilde{e}_{\Lambda}(k)\rightarrow\widetilde{e}(k)$ is uniform in
$\Lambda$ for $k$ in compact subsets of $(0,\pi\rho]$, i.e., not
containing the origin. Thus, by (\ref{16a}) and assumption {\bf i'}, for
all $k\neq 0$,   

\beqa
\nnb
\mathcal{S}_{\Lambda}(k) = \lim_{n}S_{\Lambda_n}(k_n) &\leq& 
\frac{k}{\sqrt{2}}\lim_{n}e_{\Lambda_n}(k_n)=
\frac{k}{\sqrt{2}}\lim_{n}\widetilde{e}_{\Lambda_n}(k_n)=
\frac{k}{\sqrt{2}}\lim_{n}\widetilde{e}_{\Lambda_n}(k)= \\ 
\lb{69}
&=& \frac{k}{\sqrt{2}}\widetilde{e}(k)\leq c|k|
\eeqa

\noindent where

\beq
\lb{70}
c\equiv\sup_{k\in[-\pi\rho,\pi\rho]}\Bigg\{ \frac{1}{\sqrt{2}}\sqrt{
\frac{2}{\pi\rho|k|}\mb{arccoth}\Bigg(\frac{2\pi\rho}{|k|}\Bigg)}\Bigg\} 
=\frac{1}{\sqrt{2}}\frac{1}{\pi\rho}t
\eeq

\noindent where $t\simeq 1.0481$, corresponding to the value
$k=\pi\rho$, by appendix D. This is Onsager's inequality (\ref{23}),
(\ref{24}), with $c$ given by (\ref{68}). $\blacksquare$

\noindent  {\it Remark 2}. The result of density functional theory
(appendix B) holds as an equality, only in the limit $k\rightarrow0$,
because, from (\ref{65}), (\ref{66}) and (\ref{67})  

\beq
\lb{71}
\lim_{k\rightarrow0}
\lim_{\overset{N\rightarrow\infty}{\overset{L\rightarrow\infty}
{\frac{N}{L}=\rho}}}\Bigg[\frac{1}{N}
\la B_{\Lambda}\psi_{\Lambda}^{k}|B_{\Lambda}\psi_{\Lambda}^{k}\ra
\Bigg] = \frac{1}{(\pi\rho)^2} 
\eeq 

\noindent which agrees with (\ref{b11}). However, from appendix D, we
see that (\ref{b11}) holds in very good approximation for all $k$. We
shall comment on the reasons for that a priori unexpected behaviour in
the conclusion. 

\noindent {\it Remark 3}. (\ref{64}) and (\ref{65}) show explicitly
the importance of performing the limit $k\rightarrow0$ after the
thermodynamic limit. Indeed, putting $k=0$ in (\ref{64}) we obtain a
$Pv$ on the r.h.s. which diverges. More precisely, the proof of
Proposition 2 given in appendix C becomes invalid.

\section{Conclusions and Outlook}

The Feynman variational ansatz is widely accepted today
\cite{leggett}. There is a good reason for that: when compared with
experiment, it leads to the famous phonon-roton spectrum of superfluid
${\bf ^4\text{He}}$ (see, e.g., Fig. 13.11 of \cite{huang}). The
ansatz is only expected to hold for not too large $|\vec{k}|$
($|\vec{k}|\leq k_c\approx 3.5 \mathring{A}^{-1}$), becoming unstable
with respect to the decay into several other types of excitations with
lower energies \cite{pitaevskii}. It is also well established for
trapped gases \cite{steinhauer2}.

In this paper we assumed the validity of the ansatz (assumption {\bf
iii}) and attempted to derive a precise connection between it and
Landau's criterion of superfluidity (\ref{17}). This is achieved
through the use of Onsager's inequality (\ref{23}), (\ref{24}), whose
first part was proved in section {\bf III}, Proposition 1, which
``rounds off'' the ``almost rigorous proof'' (see \cite{lieb3})
presented in \cite{price}. The second part was proved in section {\bf
IV}, Proposition 2 and Theorem 1, for Girardeau's model, a prototype
of a one-dimensional Bose fluid with pointwise repulsive
interactions. Some features of this proof - in particular the
importance of taking $k\rightarrow0$ only after the
thermodynamic limit and the role of the one-site repulsion to
guarantee the existence of the latter - will be present in higher 
dimensions. Indeed, as remarked by Lieb in \cite{lieb2}, the
fact that `` the potential is effectively a kinetic energy barrier for
large $\gamma=d/\rho$ (where $d$ is defined by (\ref{41})) - a result
that also holds in three dimensions - means that it is really
immaterial to the particles whether they can 'get around each other'
or merely 'through each other'.'' The double spectrum predicted in
\cite{lieb2} has been found experimentally \cite{steinhauer1},
substantiating the above conjectures to some extent.   

For trapped gases, a recent remarkable proof establishes superfluidity
according to one of the standard criteria \cite{lieb6}. The latter is,
however - as asserted by the authors of \cite{lieb6} - also satisfied
by the free Bose gas - in contrast to the Landau criterion. The
criteria are therefore inequivalent and it should therefore be most
interesting to prove (\ref{23}), (\ref{24}) for trapped gases, in the
limit considered in \cite{lieb5} (a review where further references
can be found) by which the range of the potential also tends to
zero. A first step is the result (\cite{lieb5}, remarks after Theorem
7.1, in particular (7.2), (7.3), (7.4)) that there is 100\%
condensation for all n-particle reduced density matrices, and
(\cite{lieb5}, Corollary 7.2) (convergence of momentum distribution
for the one-particle density matrix), but the momentum behaviour of
the two-particle density matrix remains to be studied.

It seems rather difficult to show (\ref{23}), (\ref{24})
directly for Girardeau's model: a study of correlation functions in
the model is at present restricted to Dirichlet and Neumann
b.c. \cite{forrester}. Thus, our bound for
$\mathcal{S}_{\Lambda}(\vec{k})$ (with periodic b.c.) is new even for
Girardeau's model.  

Finally, it is gratifying that our result (\ref{69}) agrees exactly
with (\ref{b11}) of density functional theory in the limit
$k\rightarrow0$, because the latter is, in a sense, like the
Thomas-Fermi theory, of semiclassical nature, which is then expected
to hold in the limit of large wave-lengths. However, Appendix D even
shows excellent agreement for all $k\in[-\pi\rho,\pi\rho]$, which may
be intuitively understood from (\ref{37}) (Proposition 1) whereby the
l.h.s. of (\ref{b11}) depends only on the \underline{variation} of
certain energies with respect to the interaction with an external
field. The latter is taken care of \underline{exactly} in density
functional theory (first term on the r.h.s. of (\ref{b3})) and is thus
a very special case, which is not affected by the problems pointed out
in \cite{lieb4} and (\cite{lieb5}, remarks after (2.12)). Since the
arguments in Appendix B are independent of the dimension, this
agreement is a very strong evidence that (\ref{23}), (\ref{24}) holds
for any dimension. 

\begin{acknowledgments}

One of us (W. W.) is very grateful to Ph. A. Martin and
V. A. Zagrebnov for fruitful discussions, and, in particular, to
Ph. A. Martin for emphasizing the importance of taking the limit
(\ref{23}) for $\vec{k}\neq\vec{0}$. We also thank the referees for
valuable critical comments. 

W. W. was partially supported by CNPq. M. A. da Silva Jr. was
supported by a grant from CNPq.

\end{acknowledgments}

\appendix

\section{}

In this appendix we study the liquid structure factor (\ref{21}) for
the free Bose gas, where

\beq
\nnb
\Omega_{\Lambda} =
\frac{1}{L^{3/2}}\otimes\cdots\otimes\frac{1}{L^{3/2}}.
\eeq

\noindent For $\vec{k}\neq\vec{0}$

\beqa
\nnb
\mathcal{S}_{\Lambda}(\vec{k}) &=&
\frac{1}{NL^{3N}}\int_{\Lambda^{N}}d\vec{x}_1\cdots d\vec{x}_N
\Bigg|\sum_{i=1}^{N}\mb{e}^{-\mb{i}\vec{k}\cdot\vec{x}_i}\Bigg|^2 \\
\nnb
&=& 1 + \frac{1}{NL^{3N}} \sum_{1\leq i\neq j\leq N}
\int_{\Lambda^{N}}d\vec{x}_1\cdots d\vec{x}_N 
\mb{e}^{-\mb{i}\vec{k}\cdot(\vec{x}_i - \vec{x}_j)} \\
\nnb
&=& 1 + \frac{N(N-1)}{NL^{3N}}L^{3(N-2)}
\int_{\Lambda^2}d\vec{x}_1 d\vec{x}_2 
\mb{e}^{-\mb{i}\vec{k}\cdot(\vec{x}_1 - \vec{x}_2)} \\
\nnb
&=& 1 + (N-1)L^{-6} \Bigg(\int_{\Lambda}d\vec{x}
\mb{e}^{-\mb{i}\vec{k}\cdot\vec{x}}\Bigg)^2. \\
\lb{a1} 
\eeqa

\noindent By (\ref{a1})

\beq
\lb{a2}
\mathcal{S}_{\Lambda}(\vec{k}) = 1 \quad\quad\mb{if}\quad\quad
\vec{k}\neq\vec{0},\quad\vec{k}\in\Lambda^{\ast}
\eeq

\noindent while

\beq
\lb{a3}
\mathcal{S}_{\Lambda}(\vec{0}) = N \quad\quad\mb{if}\quad\quad
\vec{k} = \vec{0}. 
\eeq

\noindent Thus,

\beq
\lb{a4}
\lim_{\overset{N\rightarrow\infty}{\overset{L\rightarrow\infty}{\frac{N}{L^3}=\rho}}}
\mathcal{S}_{\Lambda}(\vec{0}) = + \infty.
\eeq

\noindent (\ref{a2}) - (\ref{a4}) show that it is crucial to take the
thermodynamic limit with $\vec{k}\neq\vec{0}$, and only thereafter the
limit $\vec{k}\rightarrow\vec{0}$, in order to find the behaviour of
the liquid structure factor for large wave-length. Notice that
(\ref{a2}) must hold in order to be compatible with (\ref{20}) in the
free case.

\section{}

In this appendix we present in some detail a more precise, albeit
nonrigorous, variant of Onsager's derivation of (\ref{39}),
(\ref{40}), \cite{price}. Note that there are misprints and some
incorrections in the derivation in \cite{price}.

By the variational principle, for the Hamiltonian (\ref{38a})
((\ref{38b}) is, of course, treated the same way), the ground state
energy is 

\beq
\lb{b1}
E(N,L,\lambda) = \inf_{\psi}\la\psi|H_{\Lambda}^{(1)}(\lambda)\psi\ra.
\eeq

\noindent In (\ref{b1}), $\psi$ are normalized test functions, for $N$
Bosons in a periodic cube $\Lambda$ of side $L$. Each test function
corresponds to a test-one-particle density

\beq
\lb{b2}
\rho(\vec{x}) = N\int_{\Lambda^N}d\vec{x}_2 d\vec{x}_3\cdots
d\vec{x}_N |\psi(\vec{x},\vec{x}_2,\ldots,\vec{x}_N)|^2.
\eeq

\noindent We compute the infimum (\ref{b1}) in two steps \cite{lieb4}:
first, we \underline{fix} a test-function $\rho(\vec{x})$ and denote
by $\{\psi_{\rho}^{\alpha}\}_{\alpha}$ the class of test functions
with this $\rho$. Define the constrained energy minimum, with fixed
$\rho$, as 

\beqa
\nnb
E_{\Lambda}\{\rho\} &\equiv& 
\inf_{\alpha}\la\psi_{\rho}^{\alpha}|H_{\Lambda}^{(1)}(\lambda)
\psi_{\rho}^{\alpha}\ra \\
\lb{b3}
&=& \int_{\Lambda}v_{\lambda}(\vec{x})\rho(\vec{x})d\vec{x} +
F_{\Lambda}\{\rho\}
\eeqa

\noindent where

\beq
\lb{b4}
v_{\lambda}(\vec{x})\equiv\lambda\cos(\vec{k}\cdot\vec{x})
\eeq

\noindent and

\beq
\lb{b5}
F_{\Lambda}\{\rho\}\equiv\inf_{\alpha}\la\psi_{\rho}^{\alpha}|
H_{\Lambda}-E_0(N,L)|\psi_{\rho}^{\alpha}\ra
\eeq

\noindent is a universal functional of the density. In the second
stage, we find the infimum over all $\rho$:

\beq
\lb{b6}
E_0(N,L,\lambda) = \inf_{\rho}E_{\Lambda}\{\rho\}.
\eeq

\noindent Let $\rho_{\lambda}^{\Lambda}(\vec{x})$ correspond to the
infimum in (\ref{b6}), and $\rho^{\Lambda}(\vec{x})$ to the infimum in
(\ref{b6}) for $\lambda =0$, which we assume are attained in a
suitable space.

By the definition (\ref{28}) of $e_0$, it is reasonable to assume from
(\ref{b5}), for large $\Lambda$, that

\beq
\lb{b7}
F_{\Lambda}\{\rho\} =
\int_{\Lambda}\Bigg[e_0\{\rho_{\lambda}^{\Lambda}(\vec{x})\} - 
e_0\{\rho^{\Lambda}(\vec{x})\}\Bigg]d\vec{x}
\eeq

\noindent for some functional $e_0\{\rho\}$, such that in the
thermodynamic limit $\rho^{\Lambda}(\vec{x})\rightarrow\rho$, and
$e_0(\rho)$ is given by (\ref{28}). By (\ref{b3}), (\ref{b4}),
(\ref{b6}) and (\ref{b7}):

\beqa
\nnb
E_0(N,L,\lambda) &\equiv& \int_{\Lambda}d\vec{x}
\Bigg[e_0\{\rho_{\lambda}^{\Lambda}(\vec{x})\} - 
e_0\{\rho^{\Lambda}(\vec{x})\}\Bigg] 
+ \lambda\int_{\Lambda}d\vec{x}\rho_{\lambda}^{\Lambda}(\vec{x})
\cos(\vec{k}\cdot\vec{x}) \\
\nnb
&\cong&\frac{1}{2}e_0''(\rho)\int_{\Lambda}d\vec{x}
\Bigg[\rho_{\lambda}^{\Lambda}(\vec{x})-\rho^{\Lambda}(\vec{x})\Bigg]^2
+ \lambda\int_{\Lambda}d\vec{x}\rho_{\lambda}^{\Lambda}(\vec{x})
\cos(\vec{k}\cdot\vec{x}) \\
\lb{b8}
\eeqa

\noindent for $\Lambda$ ``sufficiently large''and $\lambda$
``sufficiently small''. Equating the functional derivative with
respect to $\rho_{\lambda}^{\Lambda}$ in (\ref{b8}) to find the
minimum, we obtain, under assumption (\ref{29}):

\beq
\nnb
e_{0}''(\rho)\Bigg[\rho_{\lambda}^{\Lambda}(\vec{x})
-\rho^{\Lambda}(\vec{x})\Bigg]
+ \lambda\cos(\vec{k}\cdot\vec{x}) =0
\eeq

\noindent and 

\beq
\lb{b9}
\rho_{\lambda}^{\Lambda}(\vec{x})-\rho^{\Lambda}(\vec{x}) =
-\frac{\lambda}{e_{0}''(\rho)}\cos(\vec{k}\cdot\vec{x}).
\eeq

\noindent Inserting (\ref{b9}) into (\ref{b8}) we find for the actual
minimum value

\beqa
\nnb
E_0(N,L,\lambda) &=& \frac{\lambda^2}{2}
(e_{0}''(\rho))^{-1}\int_{\Lambda}d\vec{x}\cos^2(\vec{k}\cdot\vec{x}) 
- \lambda^2 (e_{0}''(\rho))^{-1}\int_{\Lambda}
d\vec{x}\cos^2(\vec{k}\cdot\vec{x}) + 
\lambda\rho\int_{\Lambda}d\vec{x}\cos(\vec{k}\cdot\vec{x}) \\
\nnb
&=& -\frac{\lambda^2}{2}
(e_{0}''(\rho))^{-1}\int_{\Lambda}d\vec{x}\cos^2(\vec{k}\cdot\vec{x}) \\
\lb{b10} 
\eeqa

\noindent for $\vec{k}\neq\vec{0}$. Repeating the calculation for
(\ref{38b}), we obtain the same result as the r.h.s. of (\ref{b10})
except for the replacement of $\cos^2(\vec{k}\cdot\vec{x})$ by
$\sin^2(\vec{k}\cdot\vec{x})$. Finally, for the r.h.s. of (\ref{39})
we get

\beq
\lb{b11}
\frac{1}{N}\la B_{\Lambda}\psi_{\Lambda}^{\vec{k}}|
B_{\Lambda}\psi_{\Lambda}^{\vec{k}}\ra =
(e_{0}''(\rho))^{-1}\frac{L^3}{N}=\frac{1}{\rho e_{0}''(\rho)}
\eeq

\noindent from which (\ref{39}) and (\ref{40}) follow as an
equality. It should be clear to the reader that the above derivation,
while appealing is far from rigorous. Indeed, (\ref{b11}) does not
hold as an equality, as proved in the main text, but, as shown in
Appendix D, it is very close to the exact result. Indeed, by
(\ref{47}), the r.h.s. of (\ref{b11}) is $(1/\pi^2\rho^2)=(1/\pi^2)$
for $\rho=1$. By (\ref{65}), it must be compared with $(2/\pi)f(x)$,
where $f(x)$ is the function plotted in appendix D. The value plotted
for $x=0$ is $(1/2\pi)$, which yields the r.h.s. of (\ref{b11}), but
we see from that table that the value plotted even for $x=\pi$ is very
close to the value for $x=0$.

\section{}

\noindent {\it Proof of Proposition 2}. We must prove that the
r.h.s. of (\ref{64}) is, in the thermodynamic limit,

\beq
\lb{c1}
I\equiv\frac{8}{2\pi\rho}Pv\int_{-\pi\rho}^{\pi\rho}
\frac{dx}{k^2-4x^2} = 
\frac{8}{2\pi\rho}\,\lim_{\varepsilon\rightarrow0}
\Bigg(\int_{-\pi\rho}^{-\varepsilon}\frac{dx}{k^2-4x^2} + 
\int_{\varepsilon}^{\pi\rho}\frac{dx}{k^2-4x^2}\Bigg).
\eeq

\noindent The difficulty lies in the fact that, in the r.h.s. of
(\ref{64}), the ``integration step'' $\varepsilon_{L}\equiv
2\pi/L$ may not be directly identified with the $\varepsilon$ in
(\ref{c1}), because it depends on $L$. In Fig.1 we show the graph of the
function $f:x\rightarrow\frac{1}{(\frac{k}{2})^2-x^2}$

%%%%%%%%%%%%%%%%%%%%%%%%%%%%%%%%%%(Fig.1)%%%%%%%%%%%%%%%%%%%%%%%%%%%%%%%%%%%%%%

\begin{figure}[ht]
\psfrag{npirho}{$-\pi\rho$}
\psfrag{pirho}{$\pi\rho$}
\psfrag{A1}{$A_1$}
\psfrag{A2}{$A_2$}
\psfrag{nk/2}{$-\frac{k}{2}$}
\psfrag{k/2}{$\frac{k}{2}$}
\psfrag{passo}{$\varepsilon_{L}=\frac{2\pi}{L}$}
\psfrag{zero}{$0$}
\lb{graph}
\centerline{\epsfig{figure=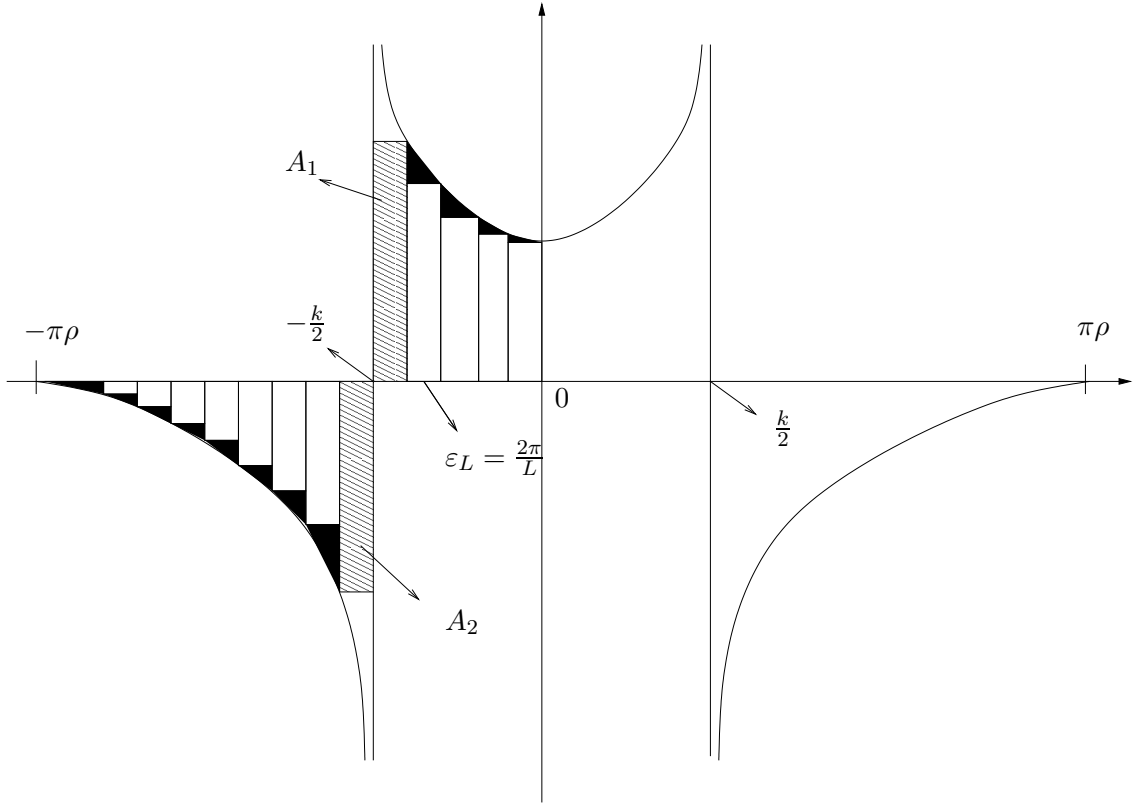,width=15cm}}
\caption{Graph of $f:x\rightarrow\frac{1}{(\frac{k}{2})^2-x^2}$}
\end{figure}

%%%%%%%%%%%%%%%%%%%%%%%%%%%%%%%%%%%%%%%%%%%%%%%%%%%%%%%%%%%%%%%%%%%%%%%%%%%%%%

\newpage

Let us call the finite sum over the $p$'s $\leq0$ on the r.h.s. of
(\ref{64}) by $\Sigma_{l}$ (the argument for the right side is the
same). 

\noindent We have 

\beq
\lb{c2}
\Sigma_{l} = A_1 + \Sigma_{l}^1 + A_2 + \Sigma_{l}^2,
\eeq

\noindent where $\Sigma_{l}^1$ is the sum of the areas of all
inscribed rectangles between $-k/2 +\varepsilon_L$ and zero, and
$\Sigma_{l}^2$ is the sum of the areas of all inscribed rectangles
between $-\pi\rho$ and $-k/2 -\varepsilon_L$.

We have, by symmetry, 

\beq
\lb{c3}
A_1 = - A_2
\eeq

\noindent while

\beq
\lb{c4}
\Sigma_{l}^1 = \int_{-\frac{k}{2}+\varepsilon_L}^{0}
\frac{dx}{(\frac{k}{2})^2-x^2}\quad +\quad \alpha_{L}^1
\eeq 

\noindent and 

\beq
\lb{c5}
\Sigma_{l}^2 = \int^{-\frac{k}{2}-\varepsilon_L}_{-\pi\rho}
\frac{dx}{(\frac{k}{2})^2-x^2}\quad -\quad \alpha_{L}^2
\eeq 

\noindent where

\beq
\lb{c6}
\alpha_{L}^1-\alpha_{L}^2
\quad\underset{L\rightarrow\infty}{\longrightarrow}0
\quad\quad\text{for fixed}\quad k\neq0. 
\eeq

\noindent In order to prove (\ref{c4}), we note that $\alpha_L^1$ is
the sum of the areas of the regions in black in Fig.1 which lie to the
left of ($-k/2-\varepsilon_L$), while $\alpha_L^2$ is the sum of the
areas in black between ($-k/2+\varepsilon_L$) and zero:

\beq
\lb{c7}
\alpha_{L}^1 =
\Bigg(\int_{-\pi\rho}^{-\frac{k}{2}-\varepsilon_L}f'(x)dx\Bigg)
\varepsilon_{L}[1+o(1/L)]
\eeq

\beq
\lb{c8}
\alpha_{L}^2 =
\Bigg(\int^{0}_{-\frac{k}{2}+\varepsilon_L}f'(x)dx\Bigg)
\varepsilon_{L}[1+o(1/L)].
\eeq

\noindent Now 

\beqa
\nnb
\varepsilon_{L}\int_{-\pi\rho}^{-\frac{k}{2}-\varepsilon_L}f'(x)dx 
&=& \varepsilon_{L}\frac{1}{(-k/2 +x)(-k/2
-x)}\Bigg|_{-\pi\rho}^{-\frac{k}{2}-\varepsilon_L} \\
\nnb
&=&\varepsilon_{L}\frac{1}{-k-\varepsilon_{L}}\frac{1}{\varepsilon_{L}}
- \varepsilon_{L}\frac{1}{(-k/2 -\pi\rho)(-k/2 +\pi\rho)} \\
\lb{c9}
&=&\frac{1}{-k-\varepsilon_{L}} + \varepsilon_{L}\frac{1}{(k/2
+\pi\rho)(\pi\rho -k/2)}
\eeqa

\noindent while

\beqa
\nnb
\varepsilon_{L}\int^{0}_{-\frac{k}{2}+\varepsilon_L}f'(x)dx &=&
\varepsilon_{L}\frac{1}{(-k/2 +x)(-k/2
-x)}\Bigg|_{-\frac{k}{2}+\varepsilon_L}^{0} \\
\lb{c10}
&=&\frac{\varepsilon_L}{(k/2)^2} + \varepsilon_{L}\frac{1}{-k
+\varepsilon_{L}}\frac{1}{\varepsilon_{L}}.
\eeqa 

\noindent For $k\neq0$, (\ref{c7}) - (\ref{c10}) establish
(\ref{c6}). Putting (\ref{c6}) into (\ref{c2}) - (\ref{c5}) we obtain
that the thermodynamic limit of the r.h.s. of (\ref{64}) indeed equals
$I$, given by (\ref{c1}). $\blacksquare$ 

\section{}

In this appendix, we show the behaviour of the function on the
r.h.s. of (\ref{66}):

\beq
\nnb
f(x)=\frac{1}{x}\text{arccoth}\Bigg(\frac{2\pi\rho}{x}\Bigg),
\eeq

\noindent defined for $x\geq0$.

Setting $\rho=1$, we construct the table:

\begin{center}
\begin{tabular}{c|c}
\hline
$x$       & $f(x)$                          \\
\hline
$0$       & \hspace{.6cm} 0.159155 ($\ast$) \\
$\pi/10$  &               0.159288          \\
$\pi/5$   &               0.159689          \\
$3\pi/10$ &               0.160365          \\
$2\pi/5$  &               0.161329          \\
$\pi/2$   &               0.162601          \\
$3\pi/5$  &               0.164205          \\
$7\pi/10$ &               0.166178          \\
$4\pi/5$  &               0.168565          \\
$9\pi/10$ &               0.171428          \\
$\pi$     &               0.174850          \\
\hline 
\end{tabular}
\end{center}

\noindent ($\ast$) is the limit (\ref{67}).

In Fig.2 below, we plot $f(x)$ for $\rho=1$:

%%%%%%%%%%%%%%%%%%%%%%%%%%%%%%%%%%(Fig.2)%%%%%%%%%%%%%%%%%%%%%%%%%%%%%%%%%%%%%%

\begin{figure}[ht]
\psfrag{x}{$x$}
\psfrag{function}{$f(x)$}
\lb{fig2}
\centerline{\epsfig{figure=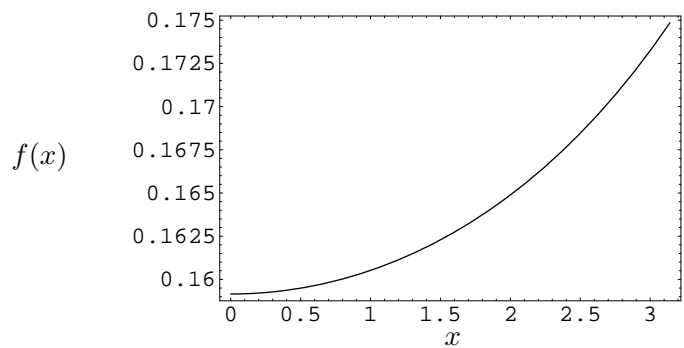,width=9cm}}
\caption{Plot of $f(x)$ for $\rho=1$}
\end{figure}

%%%%%%%%%%%%%%%%%%%%%%%%%%%%%%%%%%%%%%%%%%%%%%%%%%%%%%%%%%%%%%%%%%%%%%%%%%%%%%

We thus see that in the interval [$0,\pi$], $f$ is monotonically
increasing, and thus the largest value for $c$, given by the r.h.s. of
(\ref{68a}), is the square root of the ratio $f(\pi)/f(0)\equiv
t^2\simeq 0.174850/0.159155\simeq 1.0986$, which equals $t\simeq 1.0481$.

\bibliography{artigo}

\end{document}